\documentclass[aps,prb,10pt,twocolumn,superscriptaddress,floatfix,amsmath,amssymb]{revtex4-1}
\usepackage{hyperref}
\usepackage[usenames,dvipsnames]{color}
\usepackage[english]{babel}
\usepackage{graphicx}
\usepackage{epstopdf}
\usepackage{lineno}
\usepackage{tabularx}
\usepackage{blindtext}
\usepackage{todonotes}

\usepackage{graphicx,xcolor}
\usepackage{mathtools}
\usepackage{hyperref}
\usepackage[section]{placeins}

\begin{document}

\title {Dynamical Cluster Approximation Study of Electron Localization in the Extended Hubbard Model}

\author{Hanna Terletska}
\affiliation{Department of Physics and Astronomy, Computational Sciences Program,  Middle Tennessee State University, Murfreesboro, TN 37132, USA} 
\author{Sergei Iskakov} 
\affiliation{Department of Physics, University of Michigan, Ann Arbor, MI 48109, USA}
\author{Thomas Maier}
\affiliation{Computational Sciences and Engineering Division, Oak Ridge National Laboratory, Oak Ridge, TN 37831-6164, USA}
\author{Emanuel Gull}
\affiliation{Department of Physics, University of Michigan, Ann Arbor, MI 48109, USA}

\date{\today}
\begin{abstract}
We perform a detailed study of the phase transitions and mechanisms of electron localization in the extended Hubbard model using the dynamical cluster approximation on a $2\times 2$ cluster. We explore the interplay of charge order and Mott physics.
We find that a nearest-neighbor Coulomb interaction $V$ causes ``screening" effects close to the Mott phase transition, pushing the phase boundary to larger values of $U$. We also demonstrate the different effects of $V$ on correlations in metallic and insulating regimes, and document the different correlation aspects of charge order and Mott states.

\end{abstract}
\maketitle


\section{Introduction}

Understanding the effects of strong electron correlation and the associated  localization of charged particles remains a challenge in condensed matter physics. As various exotic quantum states including high temperature superconductivity emerge in the vicinity of localized states, metal-insulator transitions \cite{Mott,Imada_review_1998,Dagotto_2005}  are the subject of intense investigation.~\cite{Dagotto94, Basov_qm,  ANDERSON1196, Jordens_2008} 

The Hubbard model, where Coulomb interactions between electrons are assumed to be local, has been commonly used to study Mott localization and 
the related correlation-induced effects.~\cite{Hubbard_1963, Gull_HM_2021, Kivelson_2021} 
However, the approximation of purely local interactions may be severe in low-dimensional systems~\cite{Schuler_2013,Hansmann, Schuler_2013, Wehling_2011, Kotov_2012,Seo_2000, Hotta_2012, Dressel_2004, Jerome_2004, Merino_2001,Jerome_1991,Ohta_2001,Seo_2000, Friend_1999} where electron interactions are not fully screened. In these systems, non-local Coulomb interactions may lead to new physics that cannot be described by the Hubbard model. In particular, inter-site interactions are found to cause a strong modification of the effective onsite interactions resulting in a reduced Mott gap or even metallic behavior of otherwise insulating systems.~\cite{Hansmann, Ayral_2013} Moreover, they may lead to electron localization by charge ordering (CO). 

The extended Hubbard model, where  the nearest-neighbor Coulomb repulsion $V$ is included in addition to the local Hubbard interaction $U$ is a minimal model to study such effects. In it, the inclusion of $V$ energetically favors the breaking of translational symmetry with $(\pi,\pi)$ checkerboard charge ordering on a square lattice.~\cite{Zhang_1989, vanDongen_1994, Callaway_1990}

Computational tools have  played  a crucial role  in describing strong correlation physics in lattice systems. Various non-perturbative many-body methods have  been  developed.~\cite{LeBlanc15} Several of these are based on embedding schemes where the original lattice problem is mapped onto an auxiliary quantum impurity problem embedded in a self-consistently determined effective medium. The dynamical mean field theory (DMFT),~\cite{Metzner, Georges96, Vollhardt_2019} which utilizes such a mapping, has been successfully used to understand electron localization in the Hubbard model. 
Several extensions of DMFT have been developed to capture non-local spatial correlations effects.~\cite{Maier05,Kotliar_2001,Rohringer_2018,Rubtsov_2008,Rubtsov_2009,Hafermann09,Toschi_2007,Rohringer_2013,Ayral_2015,Ayral_2016,Stepanov_2019}

Similarly, quantum embedding tools have been developed for the extended Hubbard model. This includes the extended DMFT (EDMFT),~\cite{Chitra_2000,Sun_2002,Camjayi_2008,Amaricci_2010,Kapcia_2017} and non-local perturbative techniques such as EDMFT+GW~\cite{Sun_2002, Ayral_2012,Ayral_2013,Huang_2014} and the dual boson (DB) approach.~\cite{Rubtsov_2012,vanLoon_2014,Stepanov_2016,Peters_2019, Vandelli_2020}
In addition, cluster DMFT methods have been applied to explore two-dimensional (2D) extended Hubbard models on the square~\cite{Merino_2007, Terletska_2017,Merino_2007} and honeycomb lattices~\cite{WeiWu_2014} in the context of CO as well as superconductivity.~\cite{Maier_2018,Arita_2004,Merino_2001}

In this paper, we extend our previous analyses for the two-dimensional (2D) extended Hubbard model~\cite{Terletska_2017, Terletska2018,Paki} to larger values of $U$, and explore the interplay of CO and Mott physics. We construct the $V-U$ phase diagram with three different phases: the metal, the $U$-driven Mott insulator and the $V$-induced CO phase, and perform a detailed analysis of model properties upon change of $U$ and $V$. In the extended Hubbard model the electron localization emerges either via Mott localization or CO. These two ways of localization differ in behavior. For the Mott metal-insulator transition, an increase of $U$ leads to increasing correlation effects accompanied by a decrease in the double occupancy, an increase in the self-energy, and a decrease in the quasi-particle peak. For the CO phase transition, the double occupancy increases with $V$. The $V$-induced  decrease in correlations is seen in the correlated metal and insulating regimes, where the self-energy decreases with increasing $V$. Also, by exploring the properties of the CO insulating phase, we show that, unlike the Mott insulator, the CO insulator is weakly-correlated, with a band-like insulating gap opening in the spectrum.~\cite{Merino_2007} We also study the  noticeable ``screening" effects where the local on-site interaction $U$ is  effectively being reduced by non-local charge fluctuations, resulting in a shift of the Mott transition to larger values of $U$.~\cite{Hansmann}

The paper is organized as follows: in Sec.~\ref{sec: Model_and_Method} we introduce our model and briefly describe the numerical method we use in this work. In Sec.~\ref{sec:results}, we present our results. 
Sec.~\ref{sec:conclusion} contains a summary and conclusions.

\section{Model and Method}
\label{sec: Model_and_Method}
We consider the half-filled extended Hubbard model on a 2D square lattice defined by the Hamiltonian 

\begin{align}
H&=-t\sum_{\langle ij\rangle,\sigma}\left ( c_{i\sigma}^{\dagger}c_{j\sigma}+c_{j\sigma}^{\dagger}c_{i\sigma}\right )+U\sum_{i}n_{i\uparrow}n_{i\downarrow}  \nonumber \\
&+\frac{V}{2}\sum_{\langle ij\rangle,\sigma\sigma'}n_{i\sigma}n_{j\sigma'}-\mu\sum_{i\sigma}n_{i\sigma},
\label{Hamiltonian}
\end{align}
where $c_{i\sigma}(c_{i\sigma}^{\dagger})$ denotes creation (annihilation) operators of an electron with spin $\sigma=\uparrow,\downarrow$ at the lattice site $i$, $n_{i\sigma}=c^{\dagger}_{i\sigma}c_{i\sigma}$ is the number operator at site $i$; $t$ is the nearest-neighbor hopping amplitude; $U$ is the on-site interaction between electrons with opposite spins; $V$ is the inter-site interaction between two electrons on the neighboring sites, and $\mu$ denotes the chemical potential. The system is half-filled at $\mu=\frac{U}{2}+zV$, ($z$ is the coordination number). Throughout the paper we set $t=1$ as the energy unit.

In the limiting case of $V=0$, the Hamiltonian of ~Eq.\ref{Hamiltonian} describes the conventional Hubbard model with only local on-site electron-electron interaction.~\cite{Hubbard_1963} We limit our analysis to the unfrustrated Hubbard model with only nearest-neighbor hopping in the paramagnetic phase at half-filling with repulsive interactions $U>0$.

For non-zero $V$, Eq.~\ref{Hamiltonian} serves as the minimal model for describing CO induced by short-range Coulomb interactions. The emergence of CO in this model may be understood in terms of a simple energy argument: for large local interactions $(U\gg zV)$, it is energetically favorable for the systems to have a uniform  arrangement of electrons with one electron per site so that the on-site Coulomb repulsion is minimized; in the opposite limit $(zV \gg U)$, the system prefers to CO in a checkerboard arrangement of doubly occupied and empty sites, such that the off-site repulsion between electrons on nearest-neighbor sites is minimized. In mean-field approximation,~\cite{Bari_1971} a zero temperature phase transition from a Mott insulator to a CO insulator occurs at $V_c=U/z$.  Several beyond mean-field approaches have been applied to the extended Hubbard model, including the Monte Carlo on a finite size cluster,~\cite{Zhang_1989} perturbation theory, ~\cite{vanDongen_1994} variational cluster approximation,~\cite{Aichhorn_2004} the two-particle self-consistent approach,~\cite{Davoudi_2007} and, more recently, effective medium quantum embedding methods.~\cite{Chitra_2000,Bolech,Sun_2002,Camjayi_2008,Sun_2002, Ayral_2012,Ayral_2013,Huang_2014,Rubtsov_2012,vanLoon_2014,Stepanov_2016,Peters_2019, Vandelli_2020,Merino_2007, Terletska_2017,WeiWu_2014,Maier_2018,Arita_2004,Amaricci_2010,Kapcia_2017}


In our study, we solve the Hubbard ($V=0$) and the extended Hubbard ($V\neq 0$) model using the Dynamical Cluster Approximation (DCA) method, which is a momentum-space cluster extension of the DMFT.~\cite{Maier05} In the DCA, the lattice problem is mapped onto a periodic cluster of size $N_c$ embedded in a self-consistently determined dynamical effective medium.

In standard DCA for isotropic systems,~\cite{Maier05} the first Brillouin zone is divided into $N_c$ patches, each of which is denoted by a cluster momentum $K$. The lattice self-energy $\Sigma_{\sigma}(k,i\omega_n)$ within a patch is assumed to be constant and is approximated by the cluster self-energy $\Sigma_{\sigma}(K,i\omega_n)$  with $\Sigma_{\sigma}(k,i\omega_n) \approx \Sigma_{\sigma}(K,i\omega_n)$. The DCA self-consistency condition requires that at convergence, the cluster Green's function $G_{\sigma}(K,i\omega_n)$ and the coarse-grained (averaged over $N_c$ patches) lattice Green's function $\bar{G}_{\sigma}(K,i\omega_n)$ are equal. The lattice Green's function is constructed by using the cluster self-energy $\Sigma_{\sigma}(K,i\omega_n)$ and is obtained by coarse-graining as $\bar{G}_{\sigma}(K,i\omega_n)=\frac{N_c}{N}\sum_{\tilde{k}}[i\omega_n+\mu-\varepsilon (\tilde k+K)-\Sigma_{\sigma}(K,i\omega_n)]^{-1}$. Here the summation is done over the $N_c/N$ momenta $\tilde k$ within the patch about the cluster momentum $K$, with lattice momentum $k=\tilde k+K$, and $\varepsilon(k)=-2t(\cos(k_x)+\cos(k_y))$ is the lattice dispersion of the model Eq.~(\ref{Hamiltonian}) on the 2D square lattice.

The effective cluster problem is then set up using the cluster-excluded Green's function obtained via the Dyson's equation $\mathcal{G}^{-1}_{\sigma}(K,i\omega_n)=\bar{G}^{-1}_{\sigma}(K,i\omega_n)+\Sigma_{\sigma}(K,i\omega_n)$. Solving the cluster problem for a given $U$, one then gets the cluster Green's function $G_{\sigma}(K,i\omega_n)$ and cluster self-energy $\Sigma_{\sigma}(K,i\omega_n)$. The calculation is repeated iteratively until convergence is reached. For further details on the DCA and the numerical procedure, the reader is referred to Ref.~\onlinecite{Maier05}.

The major numerical work in solving the DCA self-consistency loop consists of solving the quantum cluster problem. Here we used the continuous time auxiliary field (CT-AUX)~\cite{Gull08_ctaux} method generalized to the systems with  non-local density-density inter-site interactions $V$.\cite{Terletska_2017} As described in Refs.~\onlinecite{WeiWu_2014, Arita_2004}, for square clusters, the DCA coarse-graining procedure renormalizes the nearest neighbor interactions $V$ as $\bar{V}=V \sin(\pi/N_c)/(\pi/N_c)$.

Our implementation of the DCA at finite $V$ also allows us to simulate the symmetry-broken CO phase, as long as the symmetry breaking is commensurate with the cluster.~\cite{Maier05, fuchs:2011} This enables an explicit study of not just the onset of the CO transition but also allows us
to conduct simulations directly in the CO phase.
For this, we consider a bipartite lattice structure with sublattices $A$ and $B$. To enable the CO broken-symmetry analysis, we break the translation symmetry by adding a staggered chemical potential  $\mu_i$ to our Hamiltonian of Eq. \ref{Hamiltonian}, i.e., $H_{\mu_0}=H+\sum_{i\sigma} \mu_i n_{i\sigma}$, where $\mu_i=\mu_0e^{iQr_i}$ and $Q=(\pi,\pi)$. In practice, we are interested in $\mu_o \rightarrow 0$ solution. We therefore start
the simulations with a small $\mu_0 \approx 0.1$
on the first iteration of the DCA self-consistency loop and then set $\mu_0$ to zero
on subsequent iterations. The system is then allowed to
evolve freely, and will either converge to a 
state with a uniform distribution of electron density  over lattice sites or fall into the CO state with a non-uniform charge distribution.~\cite{Terletska_2017}

Adding a staggered chemical potential $\mu_0$ breaks the translational symmetry of the lattice, leading to the doubling of the unit cell in real space. This implies that the size of the first Brillouin zone is halved, such that in the symmetry-broken CO phase the momentum space points $k$ and $k+Q$ become degenerate. Following Ref.~\onlinecite{fuchs:2011}, the broken translation symmetry introduces off-diagonal elements in DCA Green's functions $G_{\sigma}(K,K';i\omega_n)$ and self-energies $\Sigma_{\sigma}(K,K';i\omega_n)$. Consequently, the scalar DCA self-consistency equations become  $2\times2$ matrices. For example, the cluster Green's function takes on the form
    
\begin{equation}
G_\sigma(K') = 
\left( \begin{array}{cc}
    G_\sigma(K',K') & G_\sigma(K',K'+Q)  \\
    G_\sigma(K'+Q,K')  &  G_\sigma(K'+Q,K'+Q)
    \end{array} \right),
\end{equation}
where $K'$ is the momentum in the reduced Brillouin zone, and we omitted $i \omega_n$ indices. The symmetry relations for the diagonal and the off-diagonal elements of the Green's function are given as $G_{\sigma}(K',K'; i\omega_n)=-(G_{\sigma}(K'=Q; K'+Q; i\omega_n))$  and $G_\sigma(K',K'+Q)=G_{\sigma}(K'+Q,K)=G_{-\sigma}(K',K'+Q)=G_{-\sigma}(K+Q,K')$, respectively. They hold for both the Green's functions and for the self-energy. In the absence of CO, the off-diagonal elements vanish $G_\sigma(K',K'+Q)=G_\sigma(K'+Q,K')=0$ ~\cite{Terletska2018, fuchs:2011}, and the Green's function matrix becomes diagonal in momentum space. 

The linear transformation~\cite{fuchs:2011}
    \begin{align}
        G^{A/B}_{\sigma}(K') &= \frac{G_{\sigma}(K',K')+G_{\sigma}(K'+Q,K'+Q)}{2} \\ \nonumber
        &\pm G_{\sigma}(K,K+Q) 
    \end{align}
allows to study the Green's function or self-energy on each sublattice.
Here $\pm$ is used for $A$ and  $B$ sub-lattices, respectively.

\section{Results}
\label{sec:results}
%
\subsection{$V=0$ phase diagram: 
$N_c=4$ DCA Hubbard model results}
\label{sub_sect: HM}

The purpose of this section is to set the stage for the discussion of the finite $V$ extended Hubbard model in Sec.\ref{sec:V_finite}. For this, we first consider the $V=0$ paramagnetic half-filled Hubbard model on a square lattice. We focus on the temperature $T$ versus interaction strength $U$ phase diagram.


\begin{figure}[ht]
\includegraphics[width=0.45\textwidth]{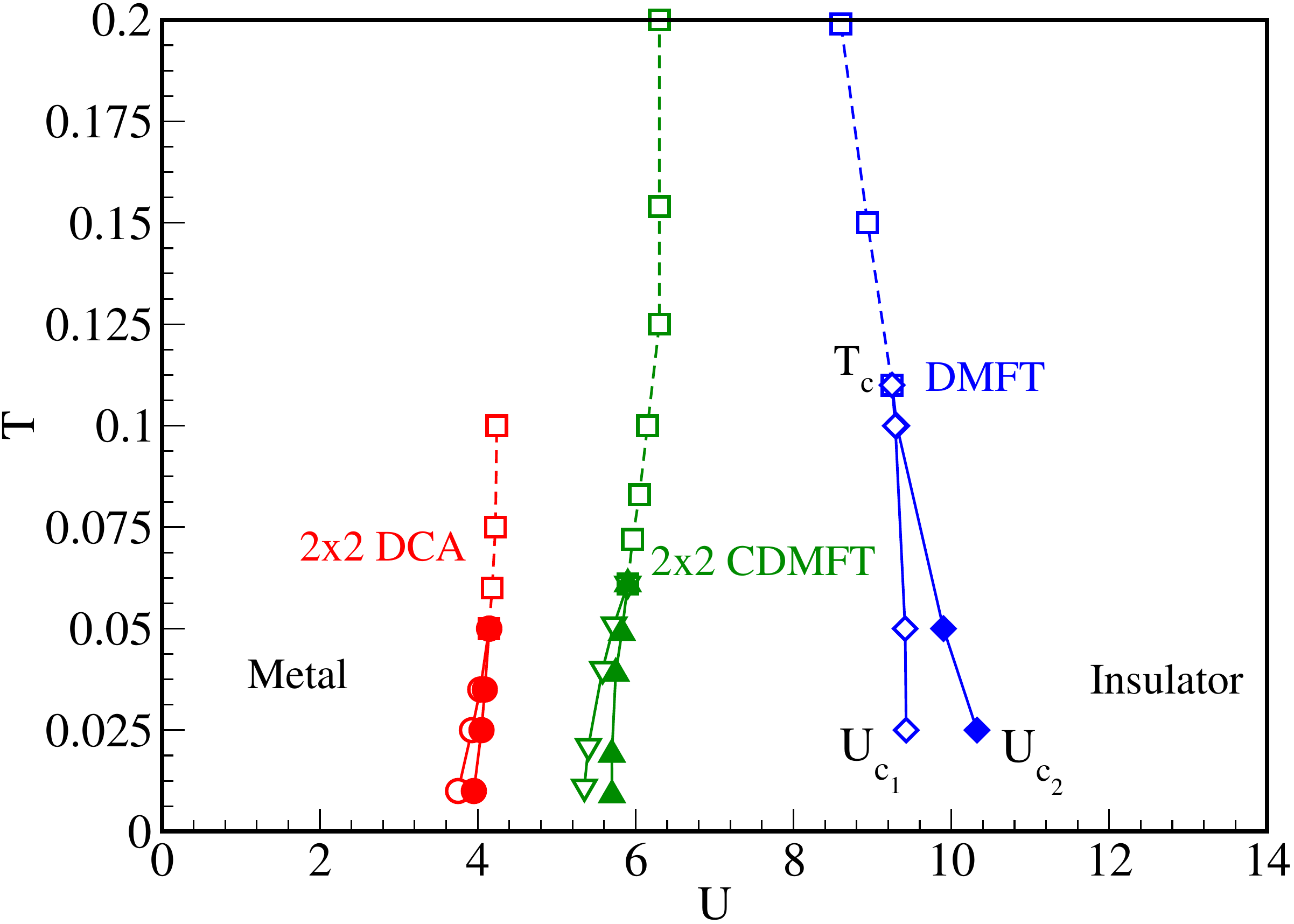}
    \caption{$T-U$ phase diagram of the paramagnetic $2D$ Hubbard model at half-filling.
  Single site DMFT (diamonds); $2\times2$ DCA (circles), $2 \times 2$ CDMFT (triangles). CDMFT data are taken from Ref.~\onlinecite{Tremblay_2019}. Closed symbols below $T_c$ mark the metallic spinodal  $U_{c_{2}}$, and open symbols mark the insulating spinodal $U_{c_1}$. Above $T_c$ the crossover between the bad metal and the bad insulator is marked by open squares. 
    }
\label{TU-pd}
\end{figure}


Fig.~\ref{TU-pd} shows the $T-U$ phase diagram for the half-filled 2D square lattice Hubbard model in the absence of the long-range order. At low temperatures, as the interaction strength $U$ increases, the system undergoes the first-order Mott-Hubbard transition between a metal and Mott insulator.~\cite{Georges96} The first-order coexistence region is delineated by two spinodal lines, $U_{c_1}$ and $U_{c_2}$. The metallic solution exists for $U<U_{c_2}$, and the insulating solution is stable for $U > U_{c_1}$. As temperature increases, the coexistence region narrows, the metallic and insulating spinodals cross at a critical point $(U_c,T_c)$ and the transition becomes continuous. The region above $T_c$ displays a crossover between metal and insulator.
Above $T_c$, various types of crossover lines have been identified depending on the criteria used.~\cite{Georges96, Rosenberg_1995, Terletska_2011_qcp, Vucicevic_2013,Vucicevic_2015,Sordi_2012,Park_2008}

For comparison purposes and to demonstrate the effect of non-local correlations beyond the DMFT, we present the results obtained by DMFT $(N_c=1)$ in addition to our $N_c=4$ DCA data. We also show the results of Ref.~\onlinecite{Tremblay_2019}  obtained by the real space CDMFT approach for a $2\times 2$ cluster.~\cite{Park_2008,Tremblay_2019}
The $N_c=4$ DCA data show that when the non-local correlations are taken into account the Mott transition remains first order, but with  significantly modified phase transition boundaries as compared to the DMFT results. These findings are in agreement with other beyond-DMFT methods, including the CDMFT of Refs.~\onlinecite{Park_2008, Tremblay_2019} and second order dual fermion results.~\cite{vanLoon_2018} This indicates that the non-local short-range anti-ferromagnetic fluctuations that are captured in the DCA significantly reduce the critical value of $U_c$ and $T_c$ at which a transition occurs.~\cite{Schafer_2015} In addition, all beyond-DMFT methods show a positive slope of the coexistence region in the phase diagram of Fig.~\ref{TU-pd}, which is different from the DMFT results. This change of slope has been explained by an entropy argument, with the low temperature insulating solution having a smaller entropy than the metal.~\cite{Park_2008}

\begin{figure}
  \includegraphics[width=0.475\textwidth]{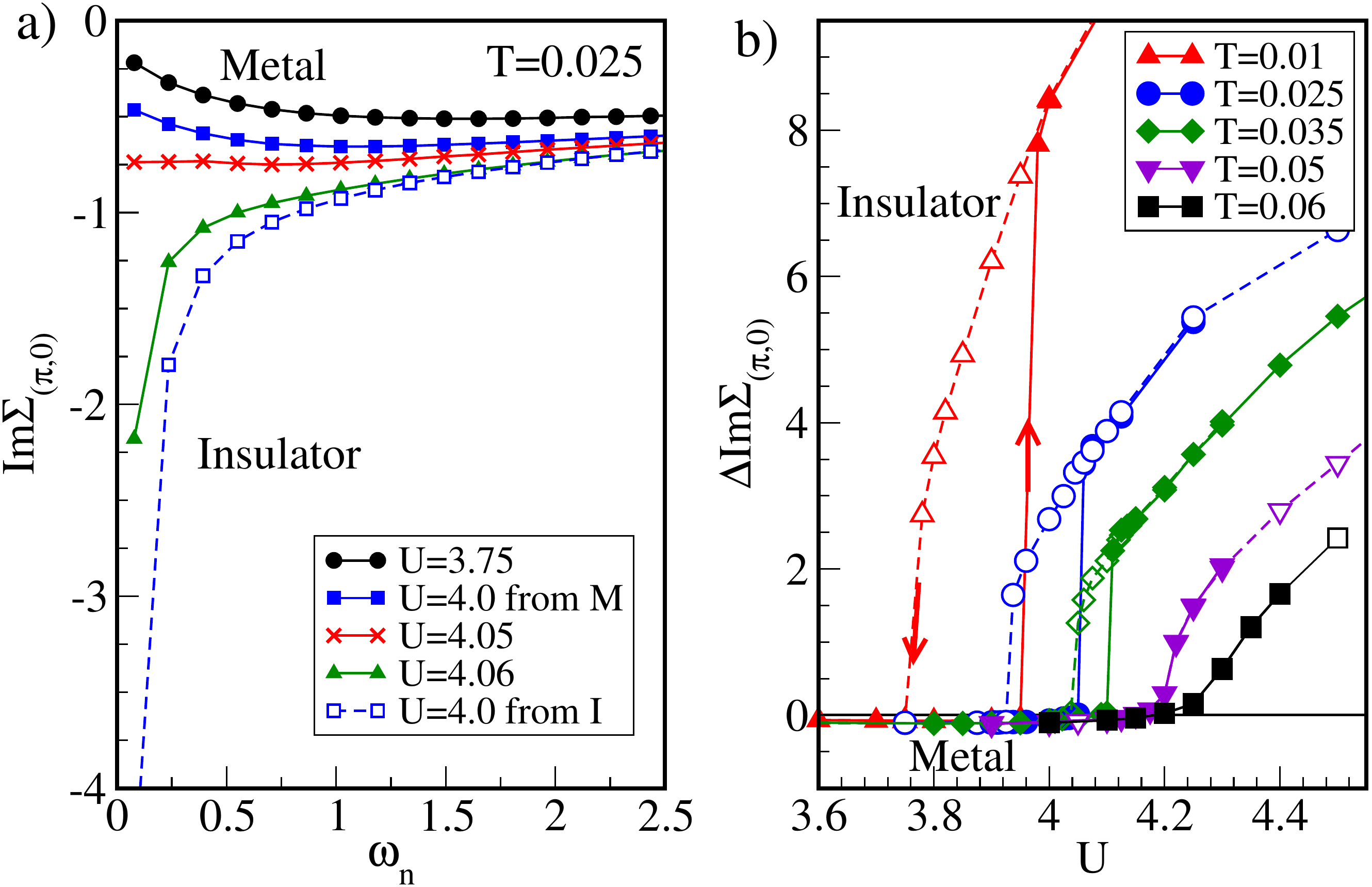}
    \caption{(a) Filled symbols: evolution of the imaginary part of the DCA self-energy $\text{Im}\Sigma_{(\pi,0)}(i\omega_n)$ for several interaction strengths $U$ obtained starting from the metallic solution at $T=0.025$. Open symbols: data for $U=4$ obtained starting from the insulating solution. (b) $\Delta \text{Im}\Sigma_{(\pi,0)}$ as a function of $U$ for indicated $T$. Hysteresis loops are obtained by sweeping the interaction strength $U$ from low to high (closed symbols) and from high to low (open symbols). Jumps in $\Delta \text{Im}\Sigma_{(\pi,0)}$ at $T<T_c$ define the spinodal points. 
    }
    \label{Fig:sigma_V0}
\end{figure}

We find that the critical value of $U_c\approx 4.15$ in $N_c=4$ DCA  ~\cite{Gull_2008}is substantially reduced compared to the DMFT value of $U_c\approx 9.35$ ~\cite{Park_2008,vanLoon_2018, Rohringer_2018} and agrees with the trend seen in the CDMFT  results,~\cite{Park_2008,Tremblay_2019} as well as with other non-local methods~\cite{Schafer_2015}, such as the VCA~\cite{Potthoff_2009_VCA} and dual fermions.~\cite{vanLoon_2018} We also find that the critical end point temperature $T_c\approx 0.05$ in $N_c=4$ DCA is substantially reduced when the non-local correlations are taken into account. Our results for $T_c$ are similar to the recent CDMFT results~\cite{Tremblay_2019} where $T_c\approx 0.06\pm0.005$ was reported. While DCA and CDMFT agree at large $N_c$,~\cite{Maier05} for smaller cluster sizes the results are expected to be different due to different embedding schemes.

All beyond DMFT methods~\cite{Schafer_2015} show that non-local correlations modify the shape of the transition lines, reduce substantially the critical values of $U_c$ for the MIT, and shrink the size of the coexistence region. 
While there is some difference in the values of $(U_c,T_c)$ for various cluster extensions of DMFT, all of the non-local methods are consistent that non-local anti-ferromagnetic correlations are strong in 2D Hubbard models and can have a dramatic effect on the Mott transition.~\cite{Schafer_2015} 

We now discuss the details of the construction of Fig.~\ref{TU-pd}. The phase transition boundaries of Fig.~\ref{TU-pd} are constructed from the analysis of the self-energy behavior presented in Fig.~\ref{Fig:sigma_V0}. To distinguish between metal and insulator, we consider the self-energy $\Sigma(K,i\omega)$ at $K=(\pi,0)$ as in Ref.~\onlinecite{Park_2008}. We confirm that for $N_c=4$ this procedure gives the same results as using the local self-energy. The Mott insulating state is identified by a divergence of the imaginary part of the self-energy $\text{Im} \Sigma_{(\pi,0)}(i\omega_n)$ at the lowest Matsubara frequency. This indicates that a pole is developed at zero frequency and the gap opens in the spectrum. We use the following metric to distinguish between metal and insulator: if $\text{Im} \Sigma_{(\pi,0)} (iw_0)>\text{Im}\Sigma_{(\pi,0)}(iw_1)$ the state is metallic, and insulating otherwise.
Fig.~\ref{Fig:sigma_V0} demonstrates such changes in the behavior from metal (e.g., at $U=3.75$) to insulator (e.g., at $U=4.06$) at $T=0.025$ as $U$ increases. The self-energy here is obtained starting from the metallic solution, and as $U$ increases,
the self-energy gets larger indicating increase of the correlations in the system, and diverges at larger $U$ values. Since $T=0.025$ is below the critical temperature $T_c$, a coexistence region exists between the metallic and insulating spinodals.  To demonstrate the coexistence of metallic and insulating solutions, we show the results for $U=4.0$ obtained starting from the metallic (solid lines) and insulating solutions (dashed lines), respectively.

To identify the spinodal $U_{c_{1}}$ and $U_{c_{2}}$ phase boundaries, based on definitions of metal and insulator, we introduce a shorthand notation $\Delta \text{Im}\Sigma_{(\pi,0)}=\text{Im}\Sigma_{(\pi,0)}(i\omega_1)-\text{Im}\Sigma_{(\pi,0)}(i\omega_0)$, which is negative for a metal and positive for an insulator. Fig. ~\ref{Fig:sigma_V0}-b) shows $\Delta \text{Im}\Sigma_{(\pi,0)}$ as function of interactions' strength $U$ at several temperatures $T=0.01, 0.025, 0.035, 0.05, 0.06$ obtained for the increasing (closed symbols) and decreasing (open symbols) values of $U$. For $T<T_c$, $\Delta \text{Im}\Sigma_{(\pi,0)}$ shows hysteresis loops, which are indications of a first-order transition. The discontinuous jumps signal the disappearance of the metallic state at $U_{c_2}$ and insulating state at $U_{c_1}$, respectively. The locations of the spinodal lines $U_{c_1}(T)$ and $U_{c_2}(T)$ are then determined by performing sweeps of $U$ at different temperatures. The hysteresis loops are widest at the lowest temperatures (e.g., $T=0.01$), decrease in size as temperature is increased ($T=0.035$), and vanish at and above the second order critical end point $T_c$ (as seen from the data for $T=0.05, 0.06$). Above the second order critical point $T_c$, the $U-$dependence of the one-particle quantities becomes smooth. This region defines the crossover region.~\cite{Georges96, Rosenberg_1995, Terletska_2011_qcp, Vucicevic_2013,Vucicevic_2015,Sordi_2012}

\begin{figure}[h!]
\includegraphics[width=0.475\textwidth]{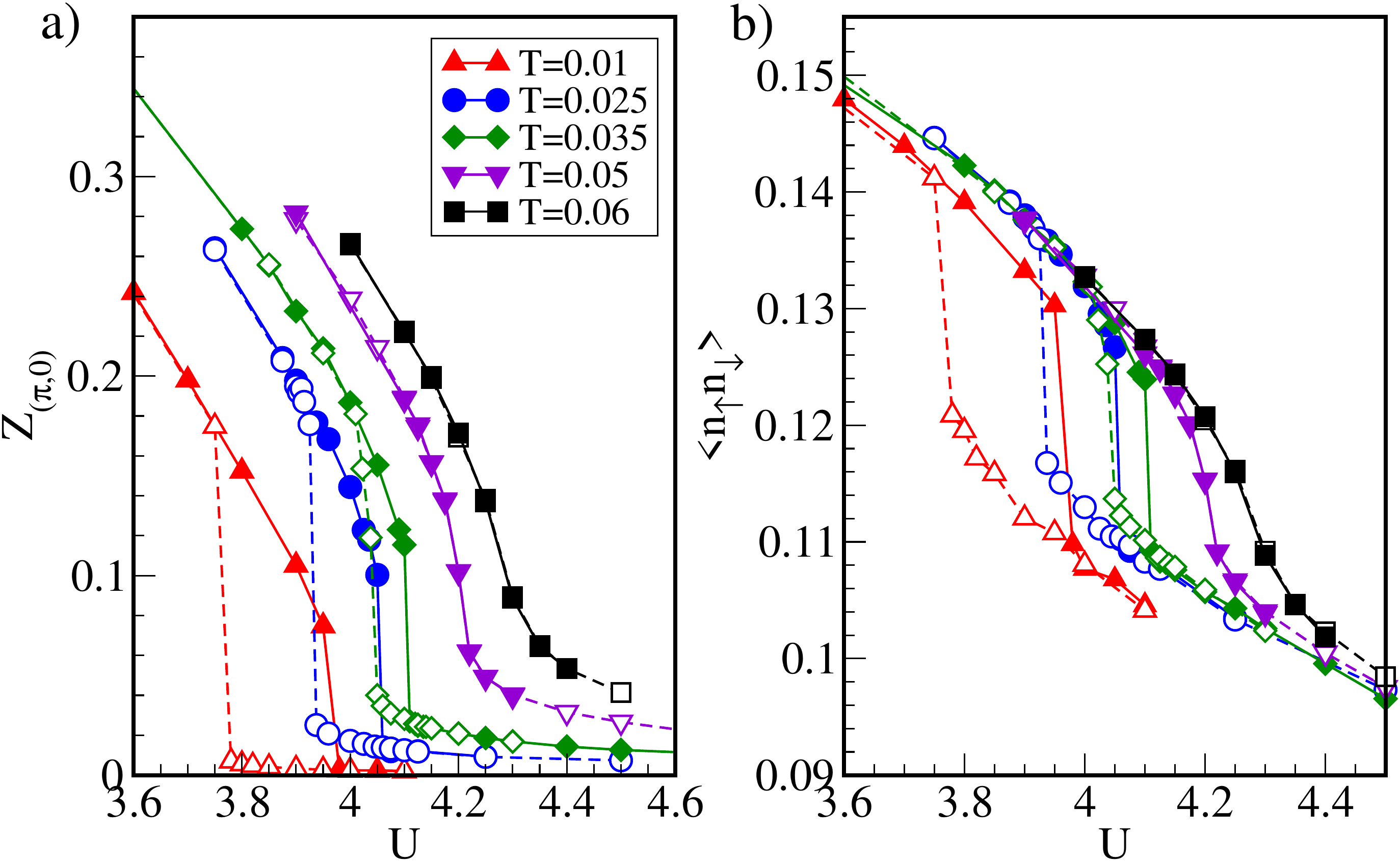}
    \caption{(a) The quasi-particle weight $Z_{(\pi,0)}$ and (b) the double occupancy $\langle n_{\uparrow}n_{\downarrow}\rangle$ as a function of interaction strength $U$ at temperatures indicated. Closed symbols mark the results obtained by starting from a metallic solution, while open symbols mark the results started from an insulating solution. }
    \label{fig: HM_Z_DOC}
\end{figure}

We show in Fig.~\ref{fig: HM_Z_DOC} the results for the quasi-particle weight $Z_{ (\pi,0)}=\lim_{i\omega_n\rightarrow 0}\left (1-\frac{\partial \text{Im}\Sigma_{K=(\pi,0)}(i\omega_n)}{\partial \omega_n} \right )^{-1}$ and the double occupancy $ \langle n_{\uparrow} n_{\downarrow} \rangle$ at several temperatures $T$ versus the interaction strength  $U$. The $U$-dependent behavior of these quantities also serves as an indicator of the correlation-induced nature of the Mott transition. We observe that the quasi-particle weight $Z_{(\pi,0)}$ gets suppressed as $U$ increases, indicating that the quasi-particles acquire a large renormalized mass as the interaction strength $U$ increases. The double-occupancy $\langle n_{\uparrow} n_{\downarrow}\rangle$ also decreases with $U$ as the Mott insulator energetically favors singly occupied over the double-occupied states. As the temperature increases, the double occupancy increases, and it takes larger values of $U$ for the Mott transition to occur. Both $Z_{(\pi,0)}$ and $\langle n_{\uparrow} n_{\downarrow}\rangle$ demonstrate hysteresis behavior at $T<T_c$, and can also be used to extract the phase boundary of the phase diagram from Fig.~\ref{TU-pd}.~\cite{Tremblay_2019}

\subsection{$V\neq 0$ phase diagram: $N_c=4$ DCA extended Hubbard model results}
\label{sec:V_finite}

\begin{figure}[ht]
  \includegraphics[width=0.475\textwidth]{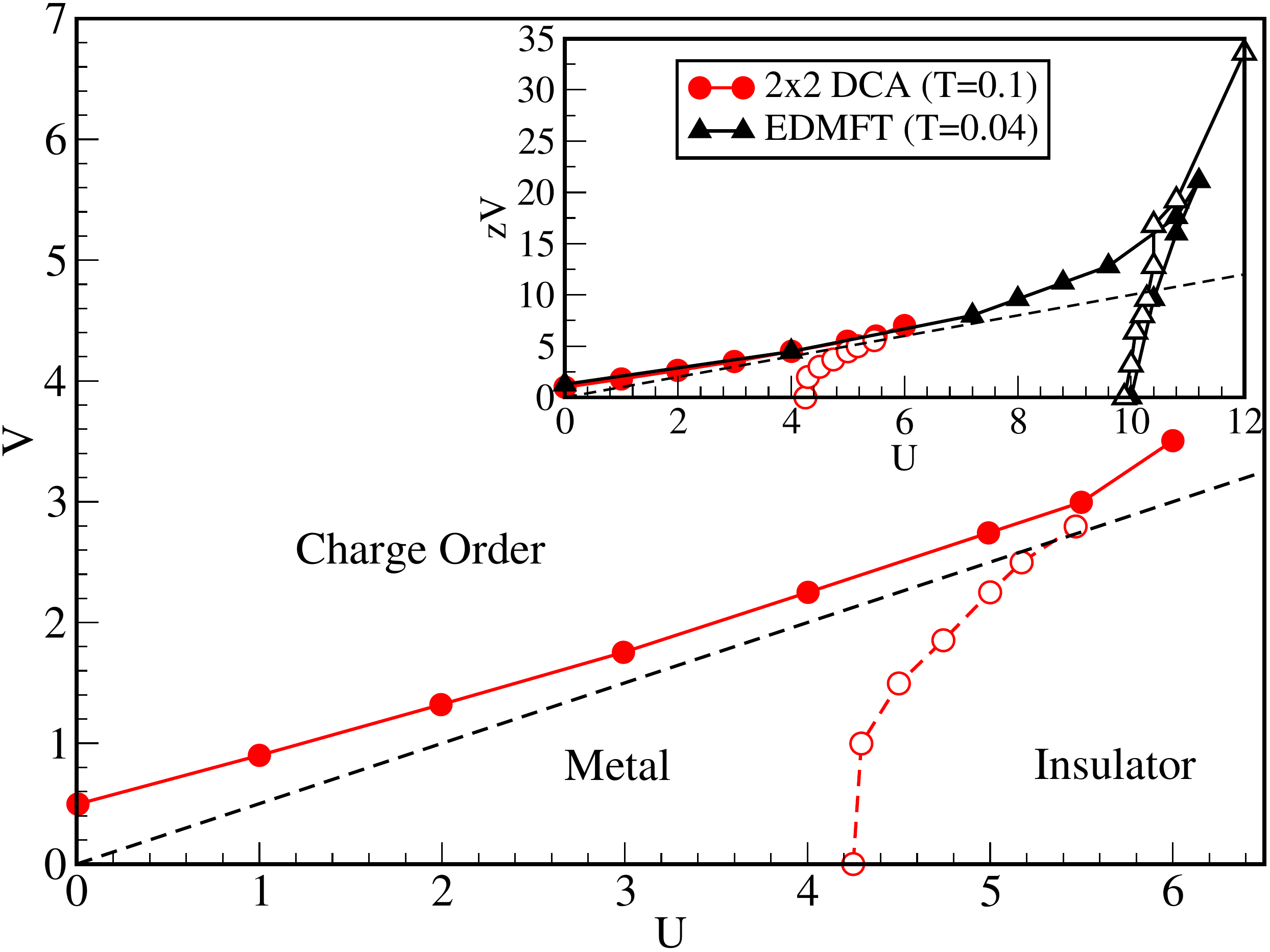}
    \caption{ The $V-U$ phase diagram of the  2D extended Hubbard model on a square lattice obtained with $Nc=4$ DCA at $T=0.1$. Three different phases are found under the changes of electron-electron interactions: the isotropic metal, the Mott insulator, and charge order. The metal to Mott insulator crossover line is shown by red open symbols, and the $V-$induced charge order phase transition boundary is denoted by filled (red) circles. The mean-field charge order phase boundary with $V=U/z$ is shown by black dashed line.
Inset: circles show $zV$ rescaled $2\times 2$ DCA data at $T=0.1$; triangles show  $zV$ rescaled EDMFT data of Ref.~\onlinecite{Medvedeva_2017}. Filled triangles are the data obtained starting from the metallic solution, and open symbols are obtained starting from the insulating solution.~\cite{Medvedeva_2017}
 }
 \label{fig:V-U_pd}
\end{figure}

To understand the effects brought about by inter-site nearest-neighbor interactions $V$, we now present results for the extended Hubbard model. First, we show in Fig.~\ref{fig:V-U_pd} the phase diagram in the presence of both local on-site interaction $U$ and nearest-neighbor interactions $V$ obtained by $N_c=4$ DCA. Due to the sign problem, especially pronounced at larger $U$ and $V$, we limit our analysis to a temperature $T=0.1$ and $N_c=4$, and $U$ values below and not far from the Mott transition. Fig.~\ref{fig:V-U_pd} shows that the phase diagram of the $2D$ extended Hubbard model at half-filling features three phases: the isotropic metal, the Mott insulator, and a charge ordered state. Similarly to the case of $V=0$, as the strength $U$ increases, the crossover from metal to Mott insulator occurs. However, with increasing $V$, the metal-insulator crossover boundary (shown by red open symbols, dashed line) occurs at larger values of $U$ due to the ``screening" effects induced by a competition between $U$ and $V$.~\cite{Hansmann,Schuler_2013,vanLoon_2014} In addition to the isotropic metallic and Mott insulating phases, the extended Hubbard model exhibits a new $V$-induced CO phase. The corresponding phase boundary (shown by filled circles) separates the parameter space  where symmetry is broken and the checkerboard arrangement of electrons with nonequivalent electron density on two sub-lattices $A$ and $B$ is energetically favorable. The values of $V$ at which the system favors charge ordering increases with increasing local interactions $U$. The DCA-determined phase boundary for the CO phase transition (obtained from the staggered density $\delta n=|n_A-n_B|$) is found to be higher than the mean-field prediction $V=U/z$ ~\cite{Bari_1971}(shown by the dashed line). 

Our results for the $V-U$ phase diagram and the trends in behavior of the phase boundaries are in agreement with the results obtained by other methods for the 2D extended Hubbard model, including ED,~\cite{Medvedeva_2017} EDMFT-based,~\cite{Sun_2002,Huang_2014,Ayral_2013} and DB~\cite{vanLoon_2014} methods. In particular, the previous studies also find that the Mott metal-insulator crossover line at finite $V$ has a positive slope, indicating that the crossover occurs at larger values of $U$.~\cite{Hansmann}
In addition, Refs.~\onlinecite{Sun_2002,Huang_2014,Ayral_2013,vanLoon_2014, Medvedeva_2017} also find that the CO boundary is above the mean field solution. The major difference with our findings is $V=0$ metal-insulator Mott crossover boundary position. We find smaller values of $U$ compared to the methods where EDMFT is a starting point. This discrepancy is due to the DMFT starting point that substantially overestimates the critical value of $U$ for the Mott transition. 
The inset of Fig.~\ref{fig:V-U_pd} shows a comparison between $2\times2$ DCA and EDMFT~\cite{Medvedeva_2017}. We note that the $2\times 2$ DCA cluster has only two distinct nearest neighbors (the one to the left is identical to the one to the right), and we therefore plot data for $zV$ with $z=2$ for $2\times2$ DCA and $z=4$ for EDMFT. Finite size effects of this system are analyzed in.~\cite{Terletska_2017}
In addition to the agreement with EDMFT (after rescaling of $V$), we find that the DCA Mott transition line remains substantially below the one from EDMFT, and that the slope of the Mott line becomes much less steep, indicating a much larger regime where $V$ suppresses the Mott transition.

\subsubsection*{a) Metal to Mott insulator crossover boundary: $V$-induced effects }
\begin{figure}
  \centering
  \includegraphics[width=0.475\textwidth]{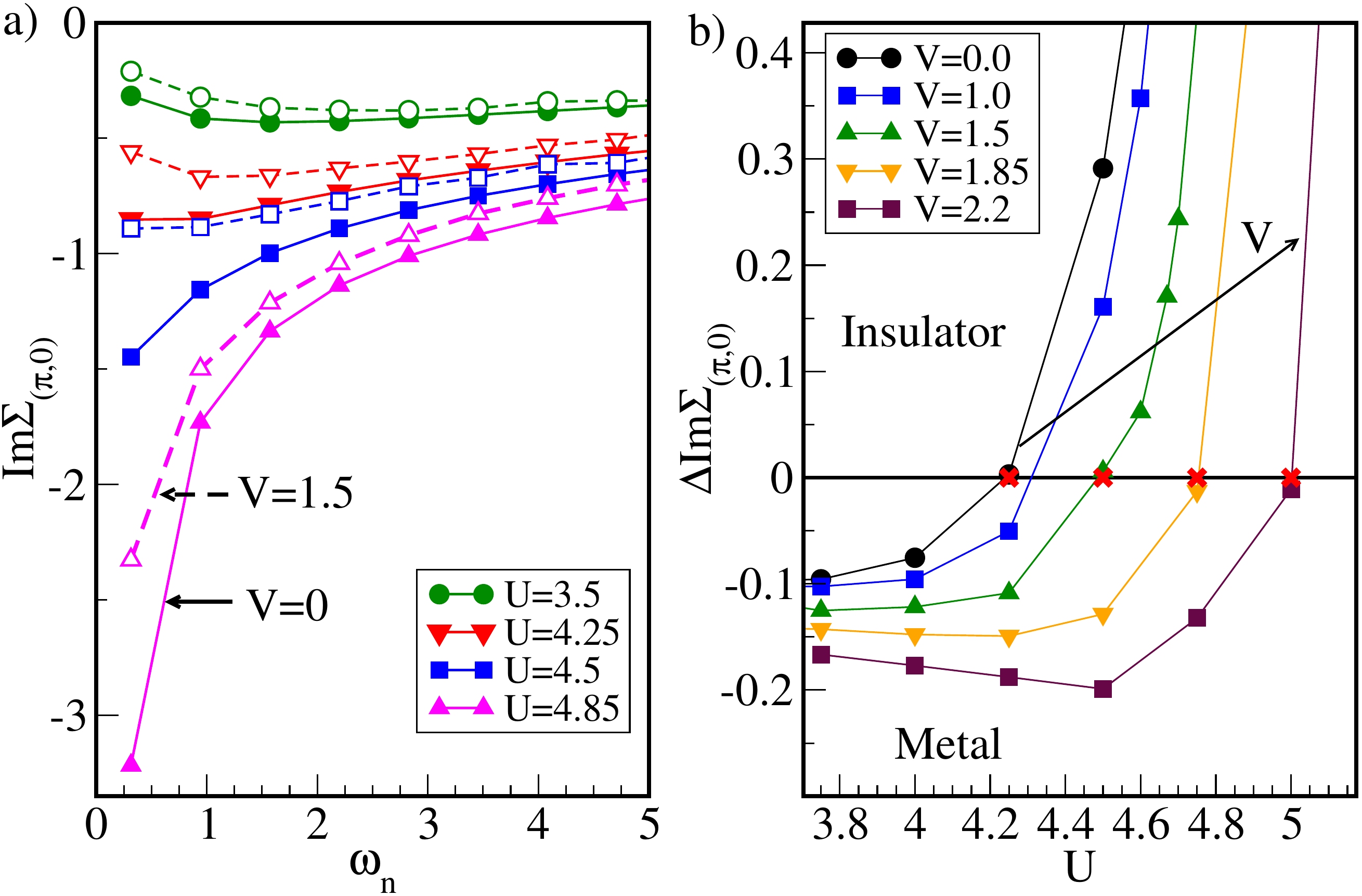}
  
   \includegraphics[width=0.475\textwidth]{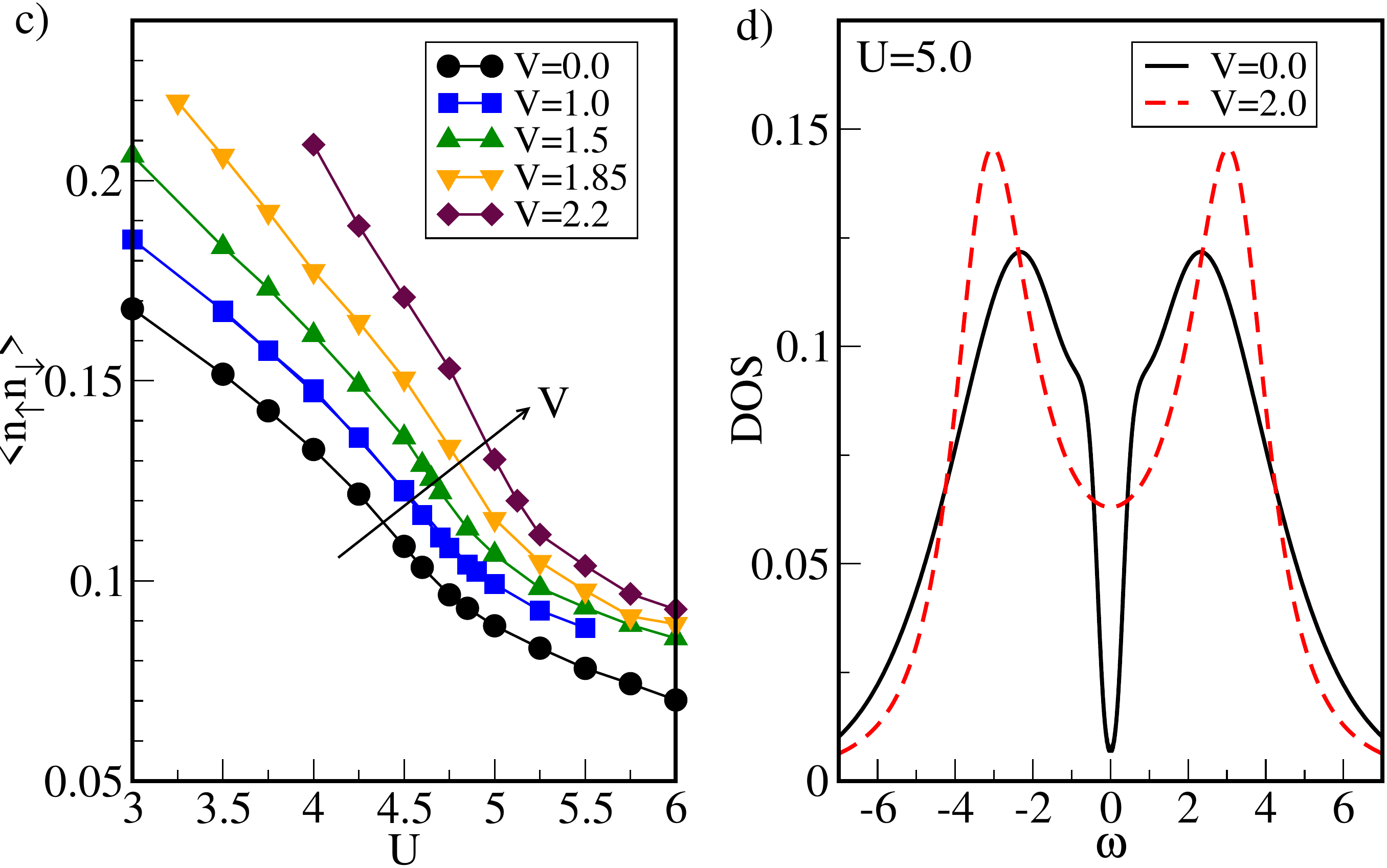}
   
    \caption{ a) The $U=3.5, 4.25, 4.5, 4.85$ and $V=0.0, 1.5$ dependencies of $\text{Im} \Sigma_{(\pi,0)}(\omega_n)$ as a function of Matsubara frequency. $V=0$: filled symbols and solid lines. $V=1.5$: open symbols and dashed lines. b) The finite $V$ metal to Mott insulator crossover boundary of Fig.~\ref{fig:V-U_pd} is determined from the change in sign of $\Delta \text{Im}\Sigma_{K=(\pi,0)}$ as function of local interactions $U$ at different values of $V$. 
    c) The double occupancy $\langle n_{\uparrow}n_{\downarrow}\rangle$ as a function of $U$ at increasing values of $V$.
    d) The $\text{DOS}(\omega)$ for $V=0$ (solid line) and $V=2.0$ (dashed line) at $U=5.0$. Other parameters: $T=0.1, N_c=4$.
}
\label{fig: V-Sigma}
\end{figure}

In this subsection, we examine more closely the effect of $V$ on the $U-$driven metal to Mott insulator crossover boundary (shown by red open symbols in Fig.~\ref{fig:V-U_pd}). 
In low-dimensional systems, large screening contributions from non-local interactions have been postulated.~\cite{Hansmann, Huang_2014,Ayral_2013} 
In fact, non-local Coulomb interactions can dramatically reduce the effective on-site interactions,~\cite{Schuler_2013} and therefore stabilize the metallic behavior against the transition to a Mott insulator.
To demonstrate this, we now examine in detail how the finite non-local interactions $V$ affect the position of the metal to Mott insulator crossover boundary of Fig.~\ref{fig:V-U_pd}. 
The position of this crossover line is determined by the same procedure we used for $V=0$ Hubbard model. The crossover points $U_c$ are determined from the change in sign of $\Delta \text{Im} \Sigma_{(\pi,0)}$ as $U$ increases, with $\Delta \text{Im}\Sigma_{(\pi,0)}<0$ for a metal, and $\Delta \text{Im}\Sigma_{(\pi,0)}>0$ for an insulator. First, to demonstrate the effect of $V$ on the self-energy, we plot in Fig.~\ref{fig: V-Sigma}-a)  $\text{Im}\Sigma_{(\pi,0)}(i\omega_n)$ as a function of Matsubara frequency for several values of $U=3.5, 4.25, 4.5, 4.85$ at $V=0$ (filled symbols) and $V=1.5$ (open symbols). As seen from Fig.~\ref{fig: V-Sigma}-a), as $U$ increases, the self-energy increases and changes behavior from metal-like ($U=3.5$) to insulator-like (U=4.5, 4.85). The metal to Mott insulator crossover occurs at $U\approx 4.25$. However, the corresponding self-energy at finite $V=1.5$ (shown by open symbols, dashed lines) is smaller compared to $V=0$. 
Consequently, at finite $V$ the crossover from the metal to insulator boundary of the 2D extended Hubbard model is pushed to larger $U$. This is also seen in Fig.~\ref{fig: V-Sigma}-b), where we plot $\Delta \text{Im}\Sigma_{(\pi,0)}$ as function of $U$ for several values of $V$. We see that the critical value of $U_c$ for which the crossover occurs $(\Delta \text{Im}\Sigma_{(\pi,0)}=0)$ increases with increasing values of the non-local interaction $V$.

The $V-$induced screening effects can also be detected in other quantities. 
In Fig.~\ref{fig: V-Sigma}-c) we plot the $U$-evolution of the double occupancy $\langle n_{\uparrow}n_{\downarrow}\rangle$ for increasing values of $V$. The inter-site interactions $V$ favor the increase of the double occupancy at a given site. As a result, it then requires larger values of $U$ to suppress the double occupancy to localize electrons in the Mott insulating phase. 
Finally, the non-local interaction driven metallicity near the Mott transition is also observed in Fig.~\ref{fig: V-Sigma}-d). Here, for comparison we plot the density of state (DOS) versus frequency $\omega$ for  $V=0$ and $V=2.0$ at $U=5.0$. We used the Pade approximation to perform the analytical continuation. Fig.~\ref{fig: V-Sigma}-d) shows that the $V=0$ insulating gap at the Fermi energy gets filled up at finite $V$, and the system becomes more metallic.

In the following, we demonstrate how non-local Coulomb interactions $V$ affect the correlations and screening in the extended Hubbard model. Ref.~\onlinecite{Schuler_2013}, using a variational principle, mapped the generalized extended Hubbard model with non-local Coulomb interactions onto an effective Hubbard model with on-site interactions being reduced according to $U^{\star}=U-\bar{V}$, where $\bar{V}$ is a weighted average of non-local interactions. 
Using this approximation, this work found that non-local Coulomb interactions, in general, can significantly weaken the local interaction effects in various low-dimensional sp-electron materials in a wide range of doping.~\cite{Schuler_2013} In particular, it has been shown that in graphene, benzene and silicene, the non-local interactions $V$-induced screening effects decrease the effective local interactions by more than factor of two, which in turn leads to stabilization of the metallic-like phase against the gapped (spin-liquid or anti-ferromagnetic Mott) insulating phases in these materials. 

\begin{figure}
  \centering
   \includegraphics[width=0.5\textwidth]{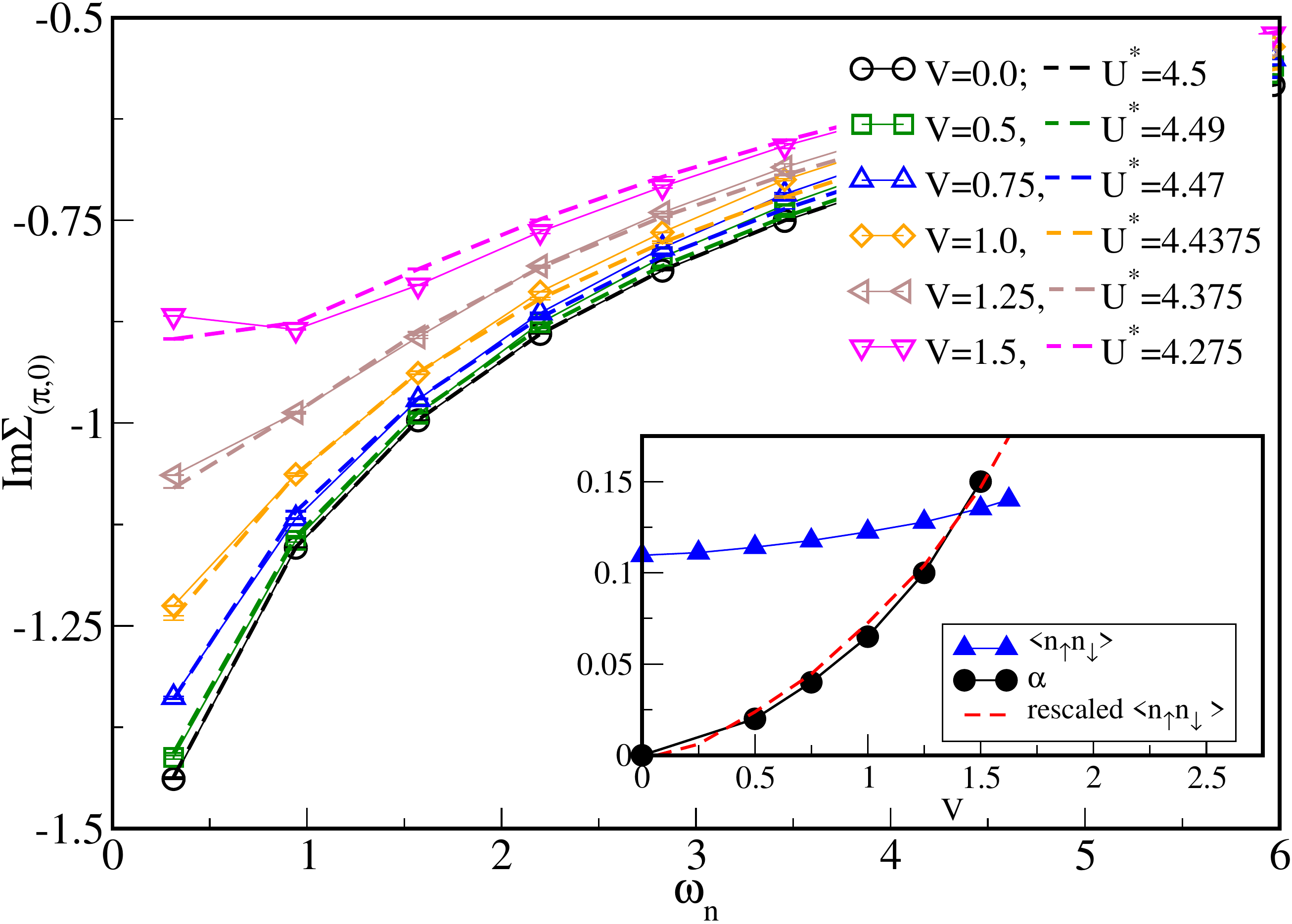}
    \caption{Imaginary part of the self-energy $Im\Sigma_{(\pi,0)}$ as function of Matsubara frequency $\omega_n$ at $K=(\pi,0)$ for different values of $V$ (open symbols) for $U=4.5$ and $T=0.1$. The data for a model with only a local screened interaction $U^{\star}$ and $V=0$ are shown by dashed lines. The inset shows the $V$ dependence of the double occupancy $\langle n_{\uparrow}n_{\downarrow}\rangle$ (blue triangles) and the pre-factor $\alpha$ (black circles) extracted from the corresponding $U^{\star}=U-\alpha V$ values. We also show the rescaled data (red dashed line) obtained as $(\langle n_{\uparrow}n_{\downarrow}\rangle_{U,V}-
   \langle n_{\uparrow}n_{\downarrow}\rangle_{V,U=0})\times A/B$, with the parameters $A,B$ being determined from fits of the original data. $\langle n_{\uparrow}n_{\downarrow}\rangle=0.11+0.0111V^2$ and $\alpha=0.06445V^2$}
   \label{fig: U_star}
\end{figure}

In Fig.~\ref{fig: U_star}, we explore this $U^\star$ behavior and demonstrate the $V$-induced screening 
in the 2D half-filled extended Hubbard model. For this we focus on the non-local interaction $V$ effects on the Mott insulating phase at $U=4.5$. First, we plot the imaginary part of the self-energy $\text{Im}\Sigma(i\omega_n)_{(\pi,0)}$ as a function of Matsubara frequency for different values of the non-local interaction $V$ (open symbols). We find that $V$ gradually reduces the self-energy, corresponding to a decrease in correlation effects. As $V$ increases, the system gradually becomes less insulating due to $V$-induced screening effects. To demonstrate further that the non-local interaction effectively reduces the local interaction $U$, we plot in Fig.~\ref{fig: U_star} the corresponding $N_c=4$ DCA  data for $U^{\star}$.
Here $U^{\star}$ is estimated by fitting the finite $V$ results for the self-energy with corresponding self-energy data obtained from $U^{\star}$ Hubbard model with $V=0$. Comparing the finite $V$ data with the $U^{\star}$ data for the self-energy in Fig.~\ref{fig: U_star}, we see that the self-energy of the extended Hubbard model with finite $V$ (we limit our analysis to $V$ below the CO phase) is well-described by the results of a Hubbard model with only local interaction $U^\star < U$. We find that the effective $U^\star$ decreases with increasing $V$.
In particular, the data shows that the non-local interaction $V$ significantly weakens the effective local interaction, according to $U^{\star}=U-\alpha V$, where $\alpha$ is a renormalized pre-factor for $V$. It was shown in Ref.~\cite{Schuler_2013} that the renormalization of the inter-site interactions $V$ can be modeled by the pre-factor related to the density-density correlation function. We analyze the $V-$dependence of the renormalized pre-factor $\alpha$, extracting it directly from our $U^{\star}$ estimates as $\alpha=(U-U^{\star})/V$. As shown in the inset of Fig.~\ref{fig: U_star}, we find that $\alpha$ and the double occupancy $\langle n_\uparrow n_\downarrow\rangle$ at a given site both increase with $V$ in a similar way (our best-fits indicate quadratic behavior with $V$, with $\langle n_{\uparrow}n_{\downarrow}\rangle=0.11+0.0111V^2$, and $\alpha=0.06445V^2$, respectively). To highlight a similar $V-$ dependence of these quantities, we also show the rescaled double occupancy data (red dashed line) obtained as $(\langle n_{\uparrow}n_{\downarrow}\rangle_{U,V}-\langle n_{\uparrow}n_{\downarrow}\rangle_{V,U=0})*A/B$, with the parameters $A,B$ being determined from the above quadratic-fits of the original data.

\subsubsection*{b) Charge order phase boundary and effect of $V$ on self-energy}
Now we focus on the CO phase boundary (Fig.~\ref{fig:V-U_pd}) 
as a function of $V$ at fixed values of $U$ and $T=0.1$. The $V-$induced CO transition is characterized by a checkerboard arrangement of electrons on the cluster sites and, hence, can be detected by a staggered density, $\delta n=n_A-n_B$, calculated in the DCA as follows:
\begin{equation}
    \delta n=\frac{2}{N_c} \left | \sum_{i \in A,\sigma} n_{i\sigma}-\sum_{i \in B,\sigma} n_{i\sigma} \right |
\end{equation}
The staggered electron density $\delta n$ describes the difference between the occupancies on the two sub-lattices $A$ and $B$ and serves as a natural order parameter for the CO phase transition, i.e., $\delta n=0$ in the uniform phase, and $\delta n \neq 0$ for the charge ordered phase.  

\begin{figure}[h]
\includegraphics[width=0.47\textwidth]{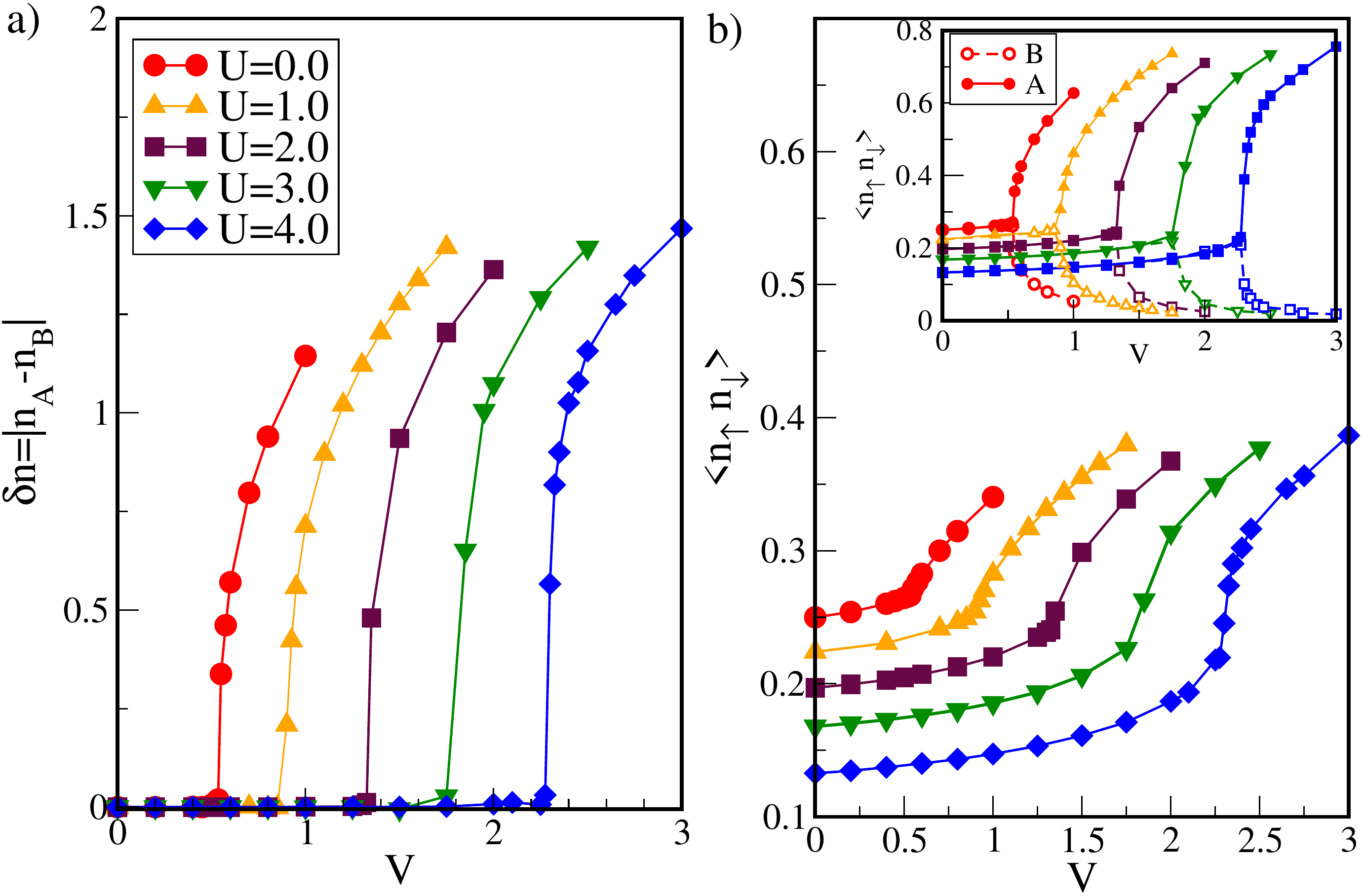}
    \caption{a) The order parameter of the CO phase $\delta n=|n_A-n_B|$ as a function of $V$ at several values of $U$. b) The double occupancy $\langle n_{\uparrow} n_{\downarrow}\rangle$ as function of $V$ for increasing values of $U$ (the same as on a) panel). The insets show the corresponding $V$-dependence of the double occupancy on the $A$ (filled symbols) and $B$ sub-lattices (open symbols). Other parameters: $T=0.1, N_c=4$.  
}
\label{fig: CO_order_arameter}
\end{figure}

In Fig.~\ref{fig: CO_order_arameter} $a)$, we show the order parameter $\delta n$ as a function of $V$ at fixed values of $U=0.0, 1.0, 2.0, 3.0, 4.0$ at fixed temperature $T=0.1$ obtained with $N_c=4$ site cluster DCA. 
At fixed $U$, increasing the non-local interaction $V$ eventually results in CO as signalled by a non-zero staggered density $\delta n \neq 0$. For larger values of $U$, the critical values of $V$ at which the transition to the CO phase occurs, increase as well. This results in the positive slope of the CO boundaries of Fig.~\ref{fig:V-U_pd}.

Similarly, the CO can be detected from the double occupancy shown in Fig.~\ref{fig: CO_order_arameter} $b)$. Notice, that in contrast to the $U-$driven Mott transition (see Fig.~\ref{fig: HM_Z_DOC}-b), where the double occupancy is suppressed with $U$), the CO transition is characterized by an overall increase of the double occupancy as $V$ increases at fixed $U$. In the inset of Fig.~\ref{fig: CO_order_arameter} $b)$, we also show the $V-$ dependence of $\langle n_{\uparrow}n_{\downarrow}\rangle$ for sub-lattices $A$ and $B$. At fixed $U$ for $V$ below the CO transition, the double occupancies $\langle n_{\uparrow}n_{\downarrow}\rangle$ on sub-lattices $A$ and $B$ are identical. Once the CO is established, the  double occupancy on the two sub-lattices become different. 

\begin{figure}
  \centering
\includegraphics[width=0.45\textwidth]{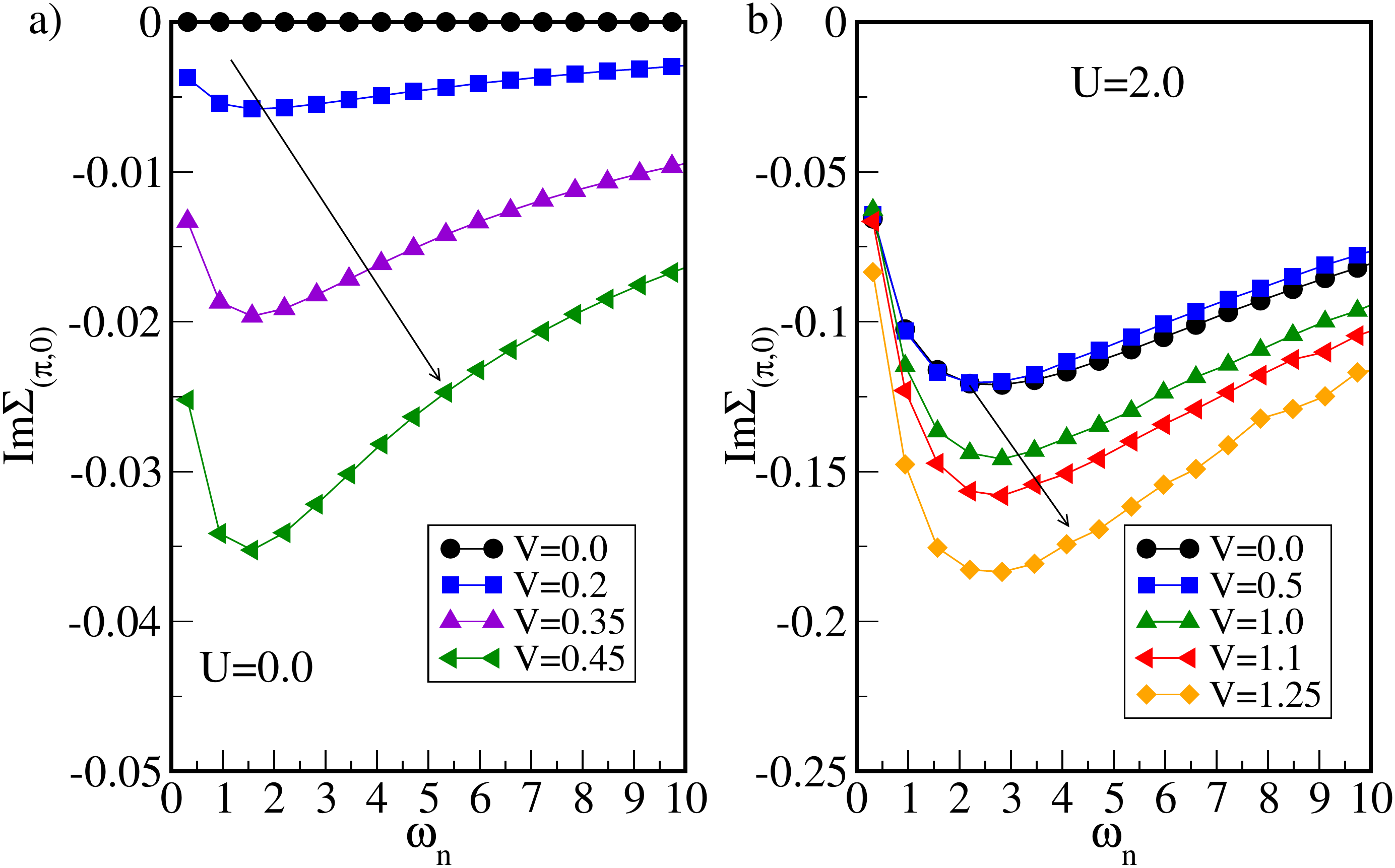}
\includegraphics[width=0.45\textwidth]{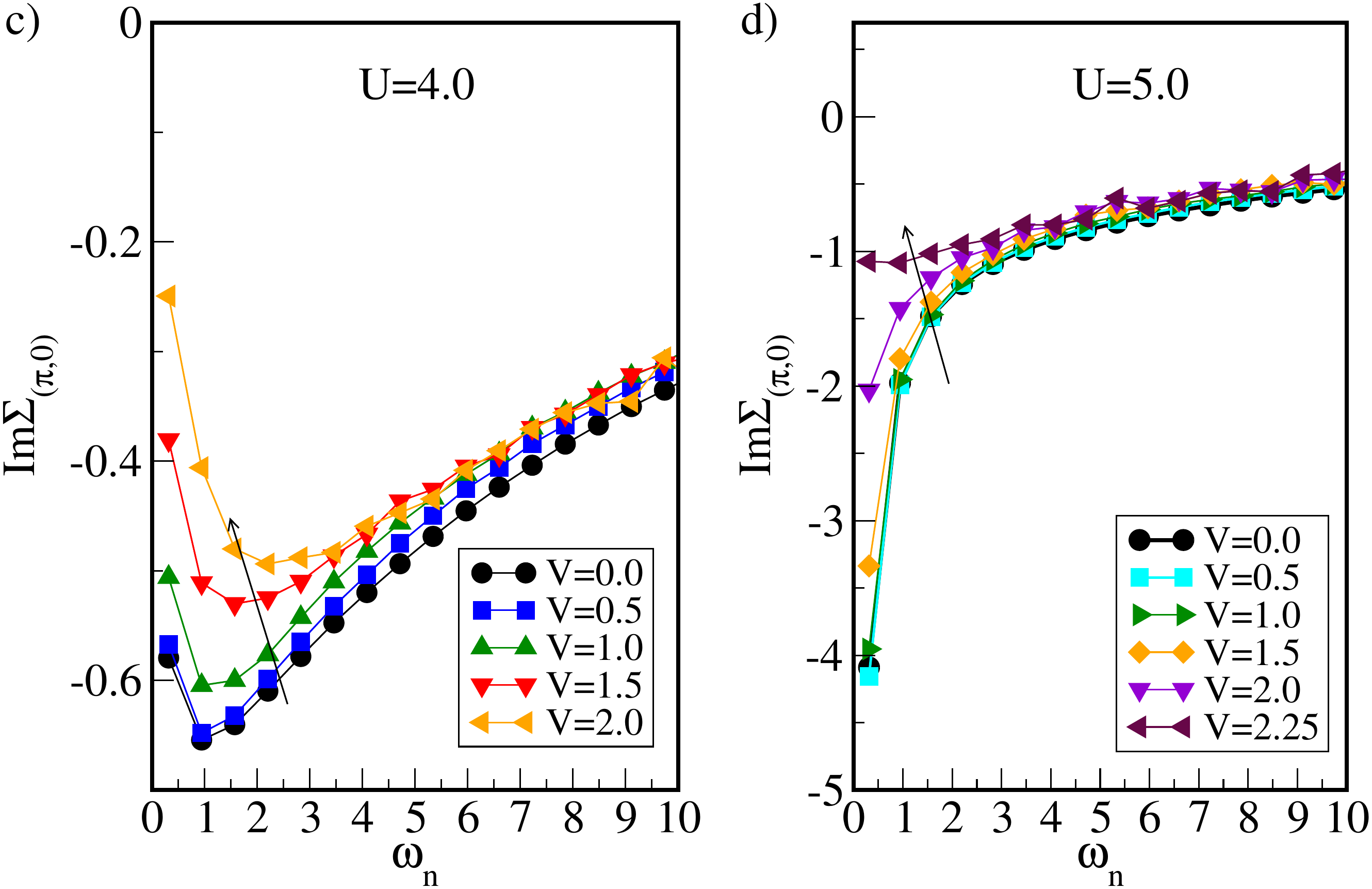}
\caption
{ 
$V$-dependence of the imaginary part of the self-energy as a function of Matsubara frequency at $K=(\pi,0)$ for small ($U\ll U_{Mott}$) and larger values of the local interaction $U \approx U_{Mott}$: a) $U=0$, b) $U=2.0$, c) $U=4.0$, d) $U=5.0$. Other parameters: $T=0.1, N_c=4$.
}
\label{fig: sigma_V}
\end{figure}

In order to further study the effects of the charge fluctuations on the single-particle dynamics, in Fig.~\ref{fig: sigma_V} we consider the effect of the inter-site interaction $V$ on the self-energy for different values of $U$ corresponding to a good metal, correlated metal, and the Mott insulator regimes. Here we plot the $V$ evolution of the imaginary part of the self-energy $\Sigma_{(\pi,0)}$ as a function of Matsubara frequency for $U=0.0, 1.0, 2.0, 3.0, 4.0$. The values of $V$ are chosen below the CO transition boundary. We begin with Fig.~\ref{fig: sigma_V} a) and b) for $U=0.0$ and $2.0$, respectively. In this case we find that as $V$ grows, $|\text{Im} \Sigma(iw_n)_{(\pi,0)}|$ increases and remains metallic. For this parameter regime, the effect of $V$ becomes very similar to the increase of the effective local interactions by $V$. A similar increase in the self-energy with $V$ for small values of $U$ has been observed in Ref.~\onlinecite{WeiWu_2014}, where for small values of $U$ the Fermi liquid behavior persisted with increasing values of $V$. In Fig.~\ref{fig: sigma_V}$ c)$ and $d)$, we show the results for larger values of $U=4.0$ and $5.0$, corresponding to the correlated metal and the Mott insulator, respectively. As $V$ increases, the magnitude of the self-energy decreases, indicating that the system becomes less insulating as a result of the screening effect of $V$.
A similar screening effect and decrease in correlations in the  presence of $V$ has been also observed in other studies of extended Hubbard models.~\cite{WeiWu_2014,Ayral_2013, Chitra_2000}

\subsubsection*{c) $U$-induced CO insulator to Metal to Mott-insulator transitions}
So far we have mainly focused on the correlation induced electron localization driven either by the non-local interaction $V$ in the CO phase, or by the local interaction $U$ in the Mott insulating case. In this subsection, we will compare these two insulating phases, and we will show that correlations can also act to induce metallic behavior. 
\begin{figure}[h!]
  \centering
\includegraphics[width=0.45\textwidth]{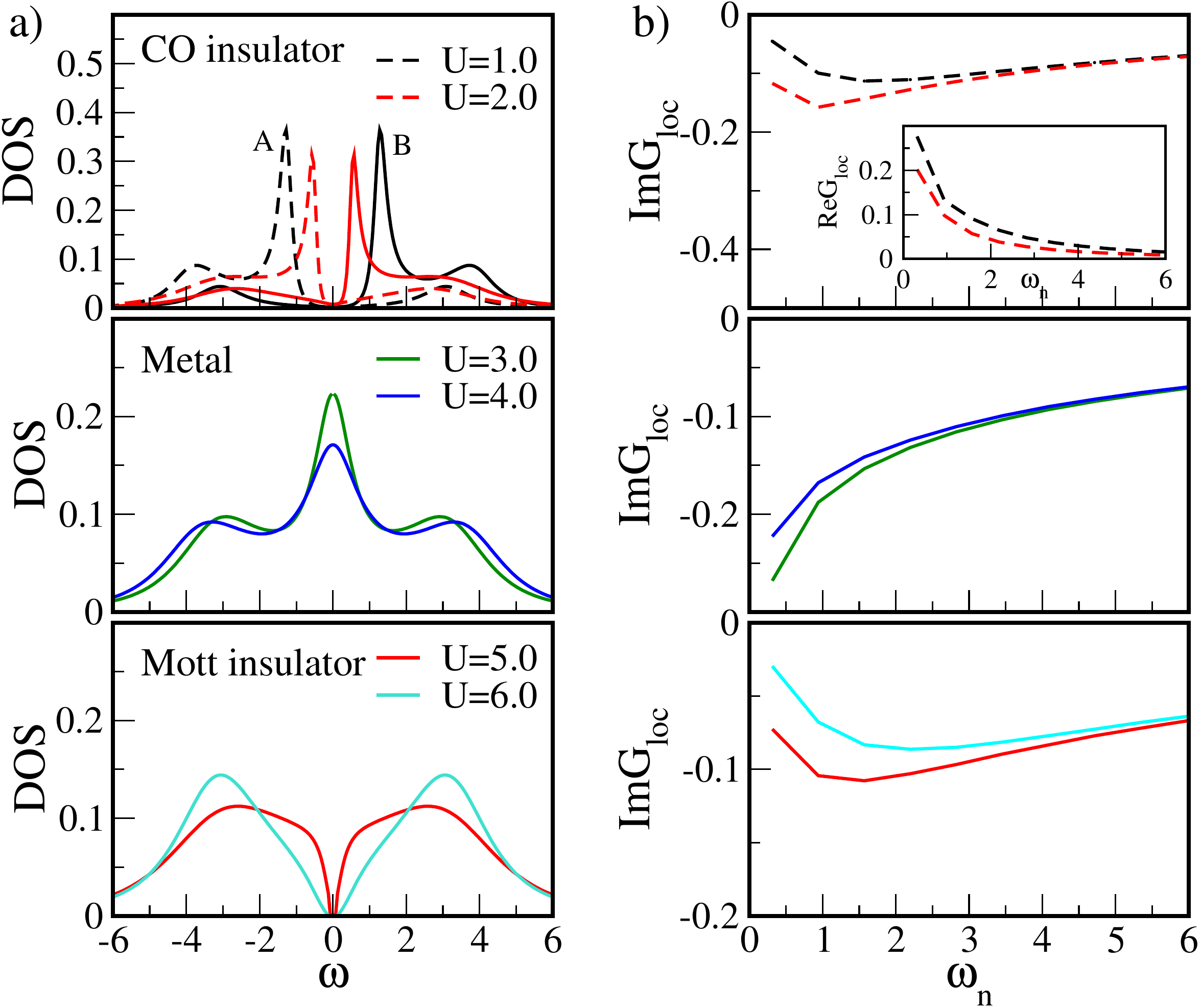}
\caption
{ Left panel: 
Density of states ($\text{DOS}$) plotted as a function of real frequency $\omega$. Data are for fixed $V=1.5$ at values of $U$ indicated. 
Right panel: the corresponding imaginary part of the local Green's function $\text{Im} G_{loc}(i\omega_n)$ as function of Matsubara frequency. 
Other parameters: $V=1.5, T=0.1, N_c=4$.
}
\label{fig: DOS_Gw}
\end{figure}

In Fig.~\ref{fig: DOS_Gw} we illustrate the correlation induced metallic behavior for the $U$-driven CO to metal, and metal to Mott insulator transitions at fixed $V=1.5$. In the left panel of Fig.~\ref{fig: DOS_Gw} we show the evolution of the DOS as the strength of $U$ changes at fixed $V=1.5$. The DOS is obtained via analytical continuation of $\text{Im} G_{loc}(w_n)$ using the Pade approximation. The data demonstrates how $\text{DOS}(\omega)$ evolves from the CO insulator to metal and the Mott insulator behavior with increasing values of $U$. At $U=1.0, 2.0$ the system is in a CO insulator state with a gap in the DOS (top panel, left column), with the $DOS_A(\omega)=DOS_B(-\omega)$. As $U$ increases, the CO gap in the DOS gets narrower, and it closes in the metallic phase (middle panel of the left column) for $U=3.0$ and $U=4.0$. Further increasing $U$ to $5.0$ and $6.0$ eventually leads to the Mott insulating behavior with an interaction-driven gap opening in the $\text{DOS}$. Such correlation-driven metallic behavior has been reported to appear in other systems featuring an electron localized insulating phase to start with. This includes systems on bipartite lattices with a staggered potential leading to a band-insulator ~\cite{Garg_2006, Wang_2020, Scalettar_2007} and  systems with disorder~\cite{Valenti_2016,Lombardo_2006}.

The corresponding Pade input data on the Matsubara axis for the imaginary part of the local Green's function $\text{Im} G_{loc}(w_n)=\frac{1}{N_c}\Sigma_{K}\text{Im}G(K,w_n)$ are shown in the right column of Fig.~\ref{fig: DOS_Gw} (b-panel). For both the CO and Mott insulators, $\text{Im}G_{loc}(w_n)$ is small and turns towards zero for $\omega_n\rightarrow 0$ consistent with the gap opening in the \text{DOS}. For the metallic case, $\text{Im}G_{loc}(w_n)$ remains finite for $\omega_n\rightarrow 0$, indicating a finite quasiparticle weight at the Fermi energy.

\begin{figure}
  \centering
\includegraphics[width=0.45\textwidth]{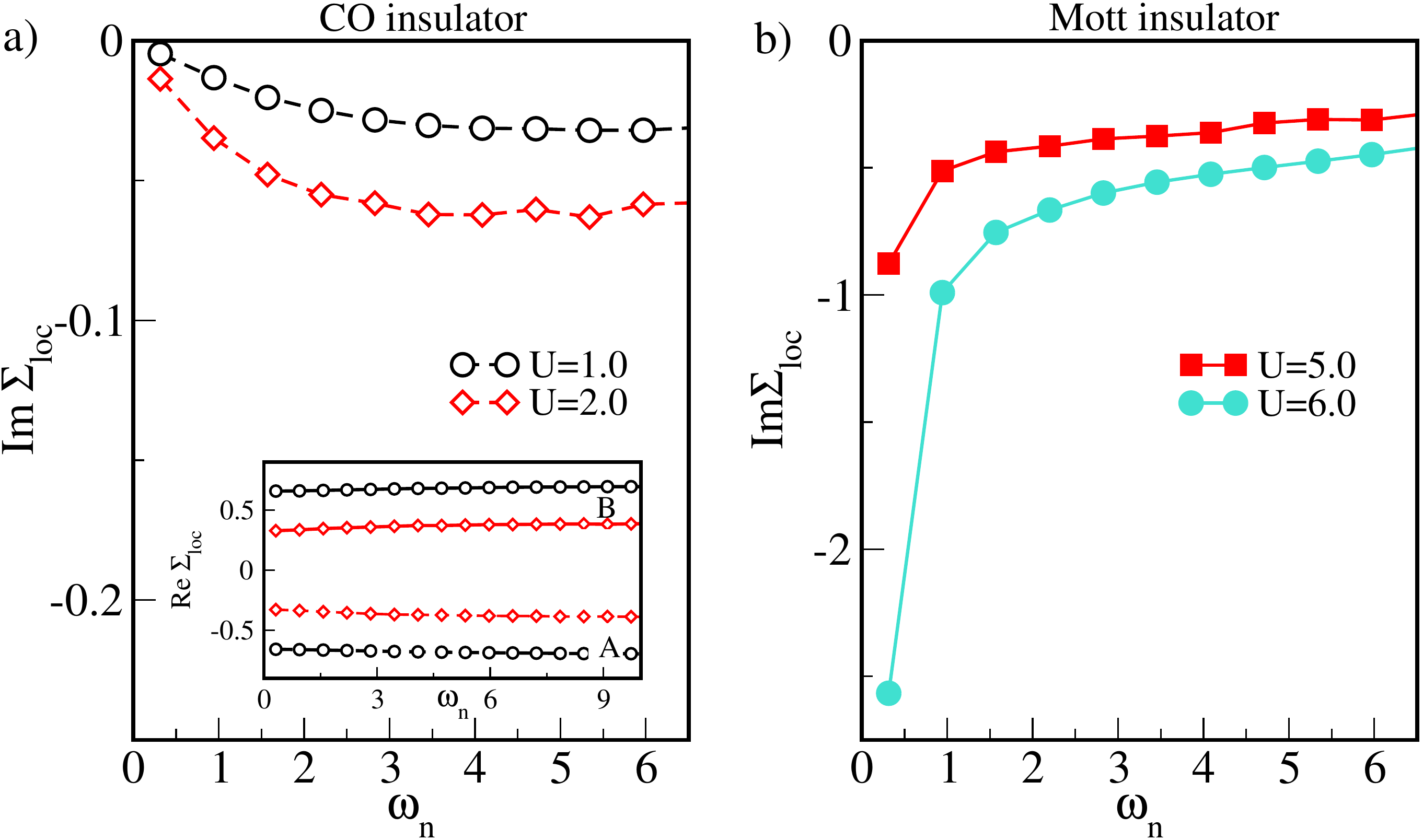}
\caption
{Imaginary part of the local self-energy $\text{Im}\Sigma _{loc}(w_n)$ for: (a) the CO insulator at $U=1.0, 2.0$; and (b) the Mott insulator at $U=5.0, 6.0$. 
Inset: real part of the local self-energy $\text{Re}\Sigma_{loc}(w_n)$ in CO state. Other parameters: $V=1.5, T=0.1, N_c=4$.
}
\label{fig: ImSigma}
\end{figure}

To further compare the CO insulator and the Mott insulator, in Fig.~\ref{fig: ImSigma} we show the Matsubara axis data for the imaginary part of the local self-energy $\text{Im}\Sigma_{loc}(w_n)$, which is a measure of the strength of correlations. The corresponding data for the local Green's function are shown in Fig.~\ref{fig: DOS_Gw}-b).  For the CO phase (Fig.~\ref{fig: ImSigma}-a), $\text{Im} \Sigma_{loc}(w_n\rightarrow 0) \rightarrow 0$ indicates that the band gap opening is due to the large $Re\Sigma_{loc}$ (shown in inset of Fig.~\ref{fig: ImSigma}-a), and the CO behaves as a weakly correlated band-insulator. In contrast, for the Mott insulator at large $U=5.0, 6.0$ (Fig.~\ref{fig: ImSigma}-b) panel), while $\text{Re} \Sigma _{loc}(w_n)=0$ (not shown), the imaginary part, $\text{Im} \Sigma_{loc}(w_n)$, is large (indicating an increased scattering rate and stronger correlations) and turns towards $-\infty$. These results clearly show the difference in the nature of the insulating CO and Mott states, characterizing the former as a band insulator and the latter as a correlation-driven Mott insulator.

\section{Conclusions}
\label{sec:conclusion}

In conclusion, using DCA on a $2\times2$ cluster, we have performed a comprehensive study of the effects of non-local correlations and interactions on the metal-insulator transitions in a 2D half-filled extended Hubbard model. We have done an analysis of the phase diagrams and the related properties for both the $V=0$ paramagnetic Hubbard model and the finite $V$ extended Hubbard model. At $V=0$, we have constructed the $T-U$ phase diagram, where we compare the DMFT, $2\times 2$ DCA, and $2\times2$ CDMFT results. We have demonstrated that in the 2D Hubbard model, the non-local correlations beyond DMFT are important; they suppress the coexistence region and significantly reduce the critical $U$ at which the transition happens, and the critical temperature $T_c$ below which the first-order transition occurs.

For the finite $V$ case, we used the DCA formalism for an extended unit cell, and constructed the $V-U$ phase diagram  for a 2D extended Hubbard model at $T=0.1$. Exploring the $V$ effects on the Mott metal-insulator crossover, we have shown that a finite nearest-neighbor interaction $V$  pushes the Mott metal-insulator crossover boundary to larger $U$ values. 
We have also demonstrated that in addition to the $U$-driven Mott localization of electrons, non-local interactions $V$ can also localize electrons via CO. We have presented a careful study of the $U$ and $V$ dependence of the order parameter, the double occupancy, self-energy and density of states. We have also shown that non-local interactions $V$ can have different effects on the self-energy behavior, depending on the values of the local interaction $U$. At larger values of $U$, the non-local interaction $V$ introduces strong screening effects with the system becoming more metallic and the self-energy mimicking the behavior of the standard $U$-only Hubbard model with a reduced effective local on-site interaction. 

To further highlight the emergence of competing states as a function of correlations $U$ and $V$ in the extended Hubbard model, we have demonstrated that in addition to localization, the electron interaction $U$ can lead to a metallic phase between the Mott and CO insulating states. Such a behavior has been argued to appear in other bipartite lattices with band-insulating phases as well as in systems with disorder.~\cite{Garg_2006, Wang_2020, Scalettar_2007,Valenti_2016,Lombardo_2006} 

Finally, comparing the $U$ and $V$-induced localization of electrons, we have shown that unlike the Mott transition, the CO transition is associated with an increase of the double occupancy and a suppression of the self-energy. The gap in the charge ordered phase is not associated with strong correlation effects, but rather with a large real part of the self-energy consistent with a band-insulator.

\acknowledgments{
This work used the Extreme Science and Engineering Discovery Environment (XSEDE), which is supported by National Science Foundation grant number ACI-1548562, through allocation DMR130036. The analysis of the results was partially conducted at the Center for Nanophase Materials Sciences, which is a DOE Office of Science User Facility. HT is supported by the National
Science Foundation under Grant No. DMR-1944974. EG acknowledges NSF DMR 2001465.
SI is sponsored by the Simons Foundation via the Simons collaboration on many-electron problem.}

\bibliography{ref.bib}
\end{document}